\documentclass[aps,prd,twocolumn,showpacs,eqsecnum,nofootinbib]{revtex4}
\usepackage[dvips]{epsfig}

\begin{document}

\title{Detection of Supernova Neutrinos by Neutrino-Proton Elastic Scattering}

\author{
\mbox{John F. Beacom$^{1}$},
\mbox{Will M. Farr$^{2}$}, 
\mbox{Petr Vogel$^{2}$}}

\affiliation{
\mbox{$^1$ NASA/Fermilab Astrophysics Center, Fermi National Accelerator 
Laboratory, Batavia, Illinois 60510-0500, USA}
\mbox{$^2$ Physics Department 161-33, Caltech, Pasadena, CA 91125 USA}
\\
{\tt beacom@fnal.gov},
{\tt farr@its.caltech.edu},
{\tt vogel@citnp.caltech.edu}}

\date{May 20, 2002}

\begin{abstract}
We propose that neutrino-proton elastic scattering, 
$\nu + p \rightarrow \nu + p$, 
can be used for the detection of supernova
neutrinos in scintillator detectors.
Though the proton recoil kinetic energy spectrum is soft,
with $T_p \simeq 2 E_\nu^2/M_p$, and the scintillation light output
from slow, heavily ionizing protons is quenched, the yield above a
realistic threshold is nearly as large as that from $\bar{\nu}_e + p
\rightarrow e^+ + n$.  In addition, the measured proton spectrum is
related to the incident neutrino spectrum, which solves a
long-standing problem of how to separately measure the total energy
and temperature of $\nu_\mu$, $\nu_\tau$, $\bar{\nu}_\mu$, and
$\bar{\nu}_\tau$.  The ability to detect this signal would give
detectors like KamLAND and Borexino a crucial and unique role in the
quest to detect supernova neutrinos.
\end{abstract}

\pacs{97.60.Bw, 13.15.+g	      \hspace{5cm} FERMILAB-Pub-02/087-A}

\maketitle

%%%%%%%%%%%%%%%%%%%%%%%%%%%%%%%%%%%%%%%%%%%%%%%%%%%%%%%%%%%%%%%%%%%%%%%%%%%%%
%			Introduction					    %
%%%%%%%%%%%%%%%%%%%%%%%%%%%%%%%%%%%%%%%%%%%%%%%%%%%%%%%%%%%%%%%%%%%%%%%%%%%%%

\section{Introduction}

When the next Galactic supernova occurs, approximately $10^4$ detected
neutrino events are expected among the several detectors around the
world.  It is widely believed that these $10^4$ events will provide
important clues to the astrophysics of the supernova as well as the
properties of the neutrinos themselves.  Interestingly, recent
breakthroughs in understanding solar and atmospheric neutrinos each
occurred when the accumulated samples of detected events first exceeded
$10^4$.

But will we have enough information to study the supernova neutrino
signal in detail?  Almost all of the detected events will be
charged-current $\bar{\nu}_e + p \rightarrow e^+ + n$, which will be
well-measured, both because of the large yield and because the
measured positron spectrum is closely related to the neutrino
spectrum. Because of the charged-lepton thresholds, the flavors
$\nu_\mu$, $\nu_\tau$, $\bar{\nu}_\mu$, and $\bar{\nu}_\tau$ can only
be detected in neutral-current reactions, of which the total yield is
expected to be approximately $10^3$ events.  However, as will be
discussed below, in general one {\it cannot} measure the neutrino
energy in neutral-current reactions.  This paper presents an exception.
These four flavors are expected
to carry away about 2/3 of the supernova binding energy, and are
expected to have a higher temperature than $\nu_e$ or $\bar{\nu}_e$.
However, there is no experimental basis for these statements, and at
present, numerical models of supernovae cannot definitively address
these issues either.  If there is no spectral signature for the
neutral-current
detection reactions, then neither the total energy carried by these
flavors nor their temperature can be separately determined from the
detected number of events.  

But it is crucial that these quantities be {\it measured}.
Both are needed for comparison to numerical supernova models.
The total energy is needed to determine the mass of the neutron star, 
and the temperature is needed for studies of neutrino oscillations.
At present, such studies would suffer from the need to
make model-dependent assumptions.  This problem has long been known,
but perhaps not widely enough appreciated.  In this paper, we clarify
this problem, and provide a realistic solution that can be implemented
in two detectors, KamLAND (already operating) 
and Borexino (to be operating soon).  The solution is
based on neutrino-proton elastic scattering, which has been observed
at accelerators at GeV energies, but has never before been shown to be
a realistic detection channel for low-energy neutrinos.  Some of our
preliminary results have been reported at
conferences~\cite{conferences}.

%%%%%%%%%%%%%%%%%%%%%%%%%%%%%%%%%%%%%%%%%%%%%%%%%%%%%%%%%%%%%%%%%%%%%%%%%%%%
%			Cross Section					   %
%%%%%%%%%%%%%%%%%%%%%%%%%%%%%%%%%%%%%%%%%%%%%%%%%%%%%%%%%%%%%%%%%%%%%%%%%%%%

\section{Cross Section}
\label{cross}

The cross section for neutrino-proton elastic scattering is an
important prediction~\cite{weinberg} of the Standard Model, and it has
been confirmed by extensive measurements at GeV energies (see, e.g.,
Ref.~\cite{ahrens}).  At the energies considered here, the full cross
section formula~\cite{weinberg,ahrens,sigma} can be greatly
simplified.  At low energies, the differential cross section as a
function of neutrino energy $E_\nu$ and struck proton recoil kinetic
energy $T_p$ (and mass $M_p$) is
\begin{eqnarray}
\label{diffxs}
\frac{d\sigma}{dT_p} & = & \frac{G_F^2 M_p}{2 \pi E_\nu^2}
\left[(c_V + c_A)^2 E_\nu^2 + 
(c_V - c_A)^2 (E_\nu - T_p)^2 \right. \nonumber \\
& - & \left. (c_V^2 - c_A^2) M_p T_p\right]\,.
\end{eqnarray}
The neutral-current coupling constants between the exchanged $Z^\circ$
and the proton are
\begin{eqnarray}
c_V & = & \frac{1 - 4\sin^2\theta_w}{2} = 0.04\,, \\ 
c_A & = & \frac{1.27}{2}\,,
\end{eqnarray}
where the factor 1.27 is determined by neutron beta decay and its
difference from unity is a consequence of the partially conserved
axial current.  Equation~(\ref{diffxs}) may be obtained directly by
summing the contributions from the valence quarks.  The cross section
for antineutrinos is obtained by the substitution $c_A \rightarrow
-c_A$.  At high energies, the primitive couplings are functions of
$q^2/M^2$, where $M \sim$ 1 GeV (the proton mass or the dipole
form-factor masses); since $q^2 = 2 M_p T_p \sim E_\nu^2$, this
variation may be safely neglected here.  At order $E_\nu/M_p$, there
is also a weak magnetism term which we have neglected.  This would
appear inside the square brackets in Eq.~(\ref{diffxs}) as
\begin{equation}
4 T_p E_\nu c_M c_A\,,
\end{equation}
where $c_M \simeq 1.4$ depends on the proton and neutron magnetic
moments~\cite{sigma}.  This term is thus positive for neutrinos and
negative for antineutrinos.  Besides being numerically small (less
than a 10\% correction), this term will cancel in the measured
differential cross section due to the indistinguishable contributions
of neutrinos and antineutrinos.  For $\nu_\mu$ and $\nu_\tau$, we
assume the same fluxes and spectra for particles and antiparticles (as
well as each other); the weak magnetism term above causes small
corrections to the emitted spectra~\cite{horowitz} that
we can neglect here.  For $\nu_e$ and
$\bar{\nu}_e$, the expected fluxes and spectra are different from each
other, but at the lower energies of these flavors the whole correction
is very small.  Other than the above points, Eq.~(\ref{diffxs}) is
correct to all orders in $E_\nu/M_p$.  As will be emphasized below,
our results are totally independent of oscillations among active flavors,
as this is a neutral-current reaction.

We use the struck proton kinetic energy in the laboratory frame as our
kinematic variable.  For a neutrino energy $E_\nu$, $T_p$ ranges
between 0 and $T_p^{max}$, where
\begin{equation}
T_p^{max} = \frac{2 E_\nu^2}{M_p + 2 E_\nu} 
\simeq \frac{2 E_\nu^2}{M_p} \,.
\end{equation}
The maximum is obtained when the neutrino recoils backwards with its
original momentum $E_\nu$, and thus the proton goes forward with
momentum $2 E_\nu$.  The other kinematic variables can be related
to $T_p$, and are
\begin{eqnarray}
\label{kinematics}
\cos\theta_p & = & 
\frac{E_\nu + M_p}{E_\nu} \sqrt{\frac{T_p}{T_p + 2 M_p}} 
\simeq \sqrt{\frac{M_p T_p}{2 E_\nu^2}} \\ 
\cos\theta_\nu & = & 
1 - \frac{M_p T_p}{E_\nu (E_\nu - T_p)}
\simeq 1 - \frac{M_p T_p}{E_\nu^2}\,,
\end{eqnarray}
where $\theta_p$ and $\theta_\nu$ are the angles of the final proton
and neutrino with respect to the direction of the incident neutrino.
In a scintillator-based detector, the proton direction cannot be
measured, so these expressions are useful just for checking the cross
section and kinematics.

\begin{figure}
\centerline{\epsfxsize=3.25in \epsfbox{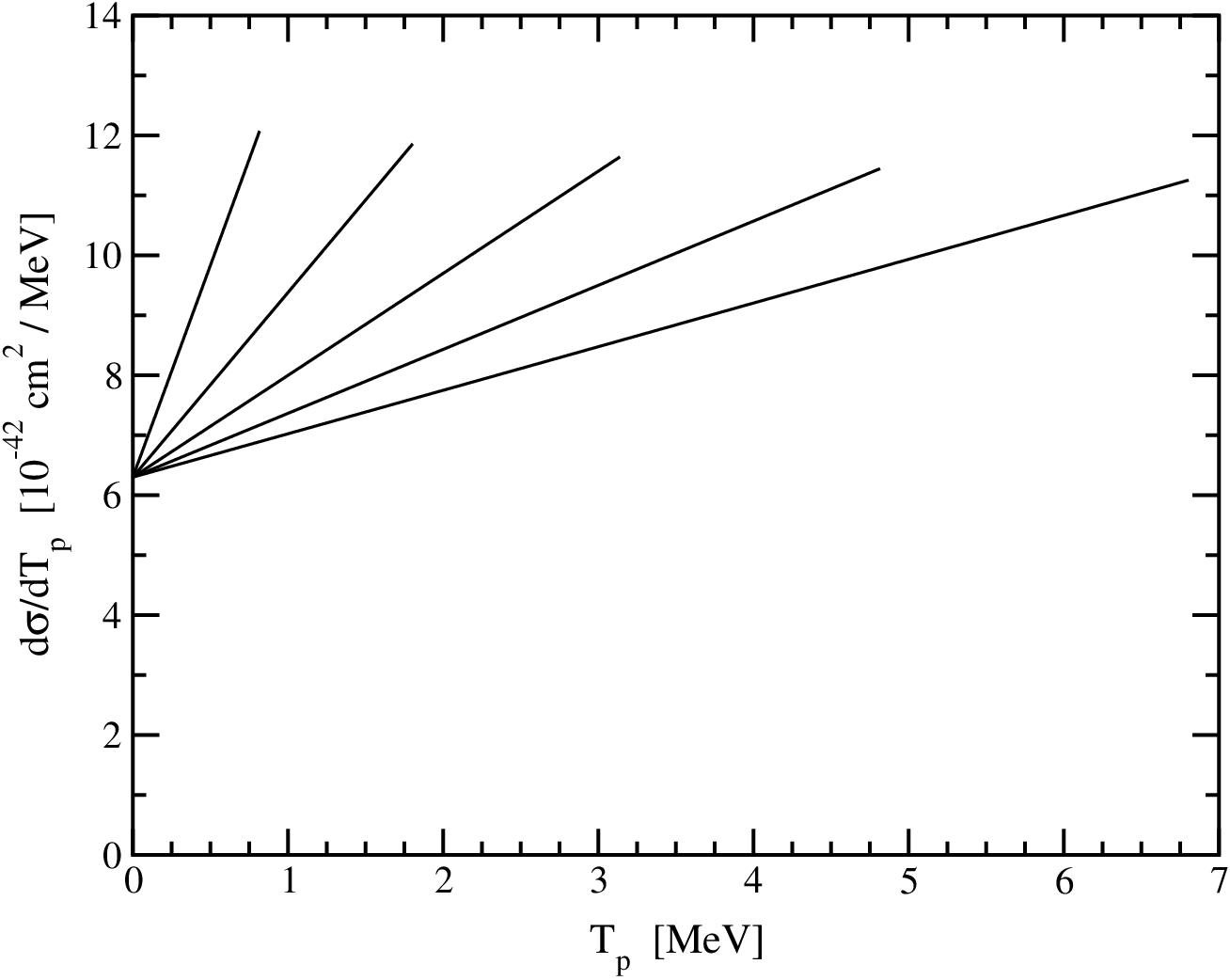}}
\caption{\label{fig:dsigdt} The differential cross section as a
function of $T_p$ for fixed $E_\nu$.  Note the rise at large $T_p$,
indicating that large kinetic energies are preferred.  From left to 
right, the lines are for $E_\nu = 20, 30, 40, 50$, and 60 MeV.}
\end{figure}

If we take $(E_\nu - T_p)^2 \simeq E_\nu^2$ (i.e., keeping only the
lowest order in $E_\nu/M_p$, a very good approximation), then the
differential cross section is very simple:
\begin{equation}
\frac{d\sigma}{dT_p} = \frac{G_F^2 M_p}{\pi} 
\left[\left(1 - \frac{M_p T_p}{2 E_\nu^2}\right) c_V^2 +
\left(1 + \frac{M_p T_p}{2 E_\nu^2}\right) c_A^2\right]\,.
\end{equation}
Since $c_A \gg c_V$, this form makes it clear that the {\it largest}
proton recoils are favored, which is optimal for detection.  Plots of
$d\sigma/dT_p$ for fixed $E_\nu$ are shown in Fig.~\ref{fig:dsigdt}.
Note that these slope in the {\it opposite} sense of the corresponding
$d\sigma/dT_e$ curves for $\nu_\mu - e^-$ scattering.  The difference
is simply due to the very different kinematics.  For neutrino-proton
elastic scattering, $T_p^{max} \simeq 2 E_\nu^2/M_p \ll E_\nu$, while for 
neutrino-electron scattering, $T_e^{max} \simeq E_\nu$.  In this
limit, the neutrino ($c_A$) and antineutrino ($-c_A$) cross
sections are identical.  If $c_V$ is neglected and the
differential cross section is expressed in terms of $\cos\theta_\nu$,
it follows the form $1 - 1/3 \cos\theta_\nu$ expected for a
non-relativistic axial coupling (i.e., a Gamow-Teller matrix element).
The total cross section is
\begin{equation}
\frac{G_F^2 E_\nu^2}{\pi} \left(c_V^2 + 3c_A^2 \right)\,.
\end{equation}
As expected, this is of the same form as the total cross section for
the charged-current reaction $\bar{\nu}_e + p \rightarrow e^+ + n$
(see, e.g., Ref.~\cite{invbeta}).  In the neutral-current case, the
vector coupling nearly vanishes, and the axial coupling is half as
large as in the charged-current channel, making the total cross section 
approximately 4 times smaller.
This factor of 4 can be immediately obtained by considering the
product of the couplings and the propagator factor, and using the
definition of $\theta_W$.

It is also interesting to compare the neutrino-proton elastic
scattering cross section with that for neutrino-electron elastic
scattering (for $\nu_\mu$ so that only the neutral-current part is
compared).  Again, the different kinematics, reflected in the maximum
kinetic energies, are crucial.  The cross section for
neutrino-electron scattering is much smaller:
\begin{equation}
\frac{\sigma_{tot}(\nu_\mu + e^-)}{\sigma_{tot}(\nu_\mu + p)} \sim
\frac{G_F^2 E_\nu m_e}{G_F^2 E_\nu^2} \sim \frac{m_e}{E_\nu},
\end{equation}
which is $\sim 10^{-2}$ for our range of energies\footnote{It is
interesting to note that the total rates of neutrino-proton elastic
scattering events from solar neutrinos are huge: in the 1 kton KamLAND
detector, the rates from the pp, $^7$Be, and $^8$B fluxes are very
roughly $10^3$/day, $10^3$/day, and $10^2$/day, respectively; however,
these are only at very low (unquenched) proton kinetic energies of
approximately 0.2 keV, 2 keV, and 200 keV, respectively.}.

%%%%%%%%%%%%%%%%%%%%%
%%%%%%%%%%%%%%%%%%%%%

In the above expressions, we have neglected contributions from strange
sea quarks in the proton~\cite{strange}.  Strange-quark effects can
enter Eq.~(\ref{diffxs}) in three ways~\cite{nustrange}.  First, the
vector form factor $c_V$ is modified by the strangeness charge radius
squared $\langle r_s \rangle^2$ by a term proportional to $q^2 \langle
r_s \rangle^2$.  Since our $q^2$ is so low, this is negligible.
Second, the magnetic form factor $c_M$ is modified by the strange
magnetic moment of the proton $\mu_s$.  This is numerically small, and
appears only in the small weak magnetism correction (see above).
Third, the
strange-quark contribution $\Delta s$ to the nucleon spin gives an
isoscalar contribution to the axial form factor $c_A$, as
\begin{equation}
c_A \rightarrow c_A = \frac{1.27}{2} - \frac{\Delta s}{2}\,.
\end{equation}
The rule for the cross section given above, of using $c_A$ for
neutrinos and $-c_A$ for antineutrinos, is also true for the
combined $c_A$ expression given here~\cite{nustrange}.
The value of $\Delta s$ is very poorly known from experiment, and is
perhaps $\Delta s = -0.15 \pm 0.15$~\cite{nustrange}.  Since $c_A
\gg c_V$, this could increase the differential cross section by
approximately 30\%, with an uncertainty of the same size.  It is
important to note that the $\Delta s$ contribution would not change 
the {\it shape} of the differential cross section, since $c_A \gg c_V$.
It may be possible to measure $\Delta s$ directly via neutrino-proton
elastic scattering at $\sim 1$ GeV in MiniBooNE~\cite{tayloe}.

%%%%%%%%%%%%%%%%%%%%%%%%%%%%%%%%%%%%%%%%%%%%%%%%%%%%%%%%%%%%%%%%%%%%%%%%%%%%%
%			Supernova Neutrinos				    %
%%%%%%%%%%%%%%%%%%%%%%%%%%%%%%%%%%%%%%%%%%%%%%%%%%%%%%%%%%%%%%%%%%%%%%%%%%%%%

\section{Supernova Neutrinos}

In this paper, we characterize the supernova neutrino signal in a very
simple way, though consistently with numerical supernova
models~\cite{SNmodels}.  The change in gravitational binding energy
between the initial stellar core and the final proto-neutron star is 
about $3 \times
10^{53}$ ergs, about $99\%$ of which is carried off by all flavors of
neutrinos and antineutrinos over about 10 s.  The emission time is
much longer than the light-crossing time of the proto-neutron star
because the neutrinos are trapped and must diffuse out, eventually
escaping with approximately Fermi-Dirac spectra characteristic of the
surface of last scattering.  In the usual model, $\nu_\mu$,
$\nu_\tau$ and their antiparticles are emitted with temperature 
$T \simeq 8$ MeV,
$\bar{\nu}_e$ has $T \simeq 5$ MeV, and $\nu_e$ has $T \simeq 3.5$
MeV.  The temperatures differ from each other because $\bar{\nu}_e$
and $\nu_e$ have charged-current opacities (in addition to the
neutral-current opacities common to all flavors), and because the
proto-neutron star has more neutrons than protons.  It is generally
assumed that each of the six types of neutrino and antineutrino
carries away about $1/6$ of the total binding energy, though this has
an uncertainty of at least $50\%$~\cite{raffeltproc}.  The supernova
rate in our Galaxy is estimated to be $(3 \pm 1)$ per century (this
is reviewed in Ref.~\cite{SNrate}).

The expected number of events (assuming a hydrogen to carbon ratio of
$2:1$) is 
\begin{eqnarray}
\label{yield}
N & = &
70.8
\left[\frac{E}{10^{53}{\rm\ erg}}\right]
\left[\frac{1{\rm\ MeV}}{T}\right] \nonumber \\ 
& \times &
\left[\frac{10{\rm\ kpc}}{D}\right]^2
\left[\frac{M_D}{1{\rm\ kton}}\right]
\left[\frac{\langle \sigma \rangle}{10^{-42}{\rm\ cm^2}}\right]\,.
\end{eqnarray}
(Though written slightly differently, this is equivalent to
the similar expression in Ref.~\cite{SNmb}.)
We assume $D = 10$ kpc, and a detector fiducial mass of 1 kton for KamLAND.
As written, Eq.~(\ref{yield}) is for the yield per flavor, assuming
that each carries away a portion $E$ of the total binding energy
(nominally, $E_B = 3 \times 10^{53}$ ergs, and $E = E_B/6$).
The thermally-averaged cross section (the integral of the cross
section with normalized Fermi-Dirac distribution) is defined for each
${\rm C H}_2$ ``molecule'', and a factor of 2 must be included for
electron or free proton targets.  
The spectrum shape of supernova
events which interact in the detector is given by the product of the
cross section and a Fermi-Dirac distribution, i.e.,
\begin{equation}
\frac{dN}{dE_\nu} \sim \sigma(E_\nu) \frac{E_\nu^2}{1 + \exp(E_\nu/T)}\,.
\end{equation}
For a cross section $\sigma \sim E_\nu^2$, this peaks at about $4 T$
(for comparison, the average neutrino energy before weighting by the
cross section is $3.15 T$), and the yield $N \sim T$.  

%%%%%%%%%%%%%%%%%%%%
\begin{figure}
\centerline{\epsfxsize=3.25in \epsfbox{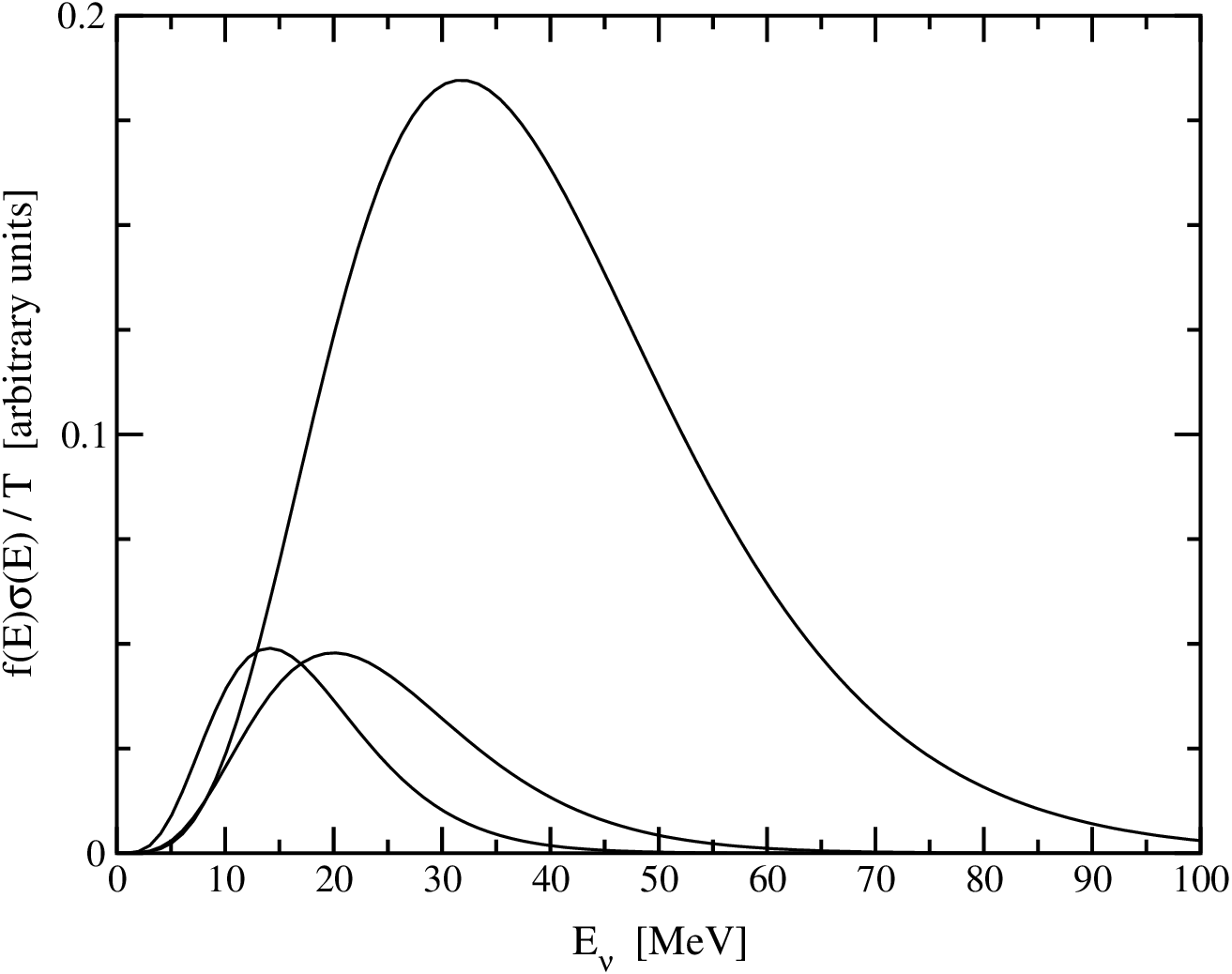}}
\caption{\label{fig:gamow} The relative spectra of neutrinos that
interact via neutrino-proton elastic scattering.  From left to right
in peak position, the curves correspond to $\nu_e$, $\bar{\nu}_e$, and
the sum of $\nu_\mu$, $\nu_\tau$, $\bar{\nu}_\mu$, and
$\bar{\nu}_\tau$.  The flux factors $N_\nu = (E_B/6)/\langle E_\nu
\rangle \sim 1/T$ have been included in the weighting.}
\end{figure}
%%%%%%%%%%%%%%%%%%%%

Prior to this paper, the largest expected yield in any oil or water
detector was from $\bar{\nu}_e + p \rightarrow e^+ + n$.  As noted in
Section~II, the total cross sections for charged-current $\bar{\nu}_e
+ p \rightarrow e^+ + n$ and neutral-current $\nu + p \rightarrow \nu
+ p$ have similar forms, though the latter is about 4 times smaller.
However, this is compensated in the yield by the contributions of all
six flavors, as well as the higher temperature assumed for $\nu_\mu$
and $\nu_\tau$ ($T = 8$ MeV instead of 5 MeV).  Thus, the total yield
from $\nu + p \rightarrow \nu + p$ is {\it larger} than that from
$\bar{\nu}_e + p \rightarrow e^+ + n$, when the detector threshold
is neglected.

Taking into account radiative, recoil, and weak magnetism corrections,
the thermally-averaged cross section for $\bar{\nu}_e + p \rightarrow
e^+ + n$ at $T = 5$ MeV is $44 \times 10^{-42}$ cm$^2$ (for 2
protons)~\cite{invbeta}.  These corrections reduce the
thermally-averaged cross section by about 20\%, and also correct the
relation $E_e = E_\nu - 1.3$ MeV.  The total expected yield from this
reaction is thus about 310 events in 1 kton.

In Fig.~\ref{fig:gamow}, the relative contributions to the spectra of
neutrinos that interact in the detector are shown.  The integral for
the combined yield from $\nu_\mu$, $\nu_\tau$, $\bar{\nu}_\mu$, and
$\bar{\nu}_\tau$ clearly dominates.  Further, since the differential
cross section favors large $T_p$, and since $T_p \sim E_\nu^2/M_p$,
the corresponding proton recoil kinetic energy spectrum will be much
harder, so that they will be even more dominant above a realistic
detector threshold.

Since the struck protons in $\nu + p \rightarrow \nu + p$ have a
relatively low-energy recoil spectrum, and since realistic detectors
have thresholds, it is crucial to consider the proton spectrum in
detail, and not just the total yield of neutrinos that interact.

%%%%%%%%%%%%%%%%%%%%%%%%%%%%%%%%%%%%%%%%%%%%%%%%%%%%%%%%%%%%%%%%%%%%%%%%%%%%%
%			Proton Spectrum					    %
%%%%%%%%%%%%%%%%%%%%%%%%%%%%%%%%%%%%%%%%%%%%%%%%%%%%%%%%%%%%%%%%%%%%%%%%%%%%%

%%%%%%%%%%%%%%%%%%%%%
\begin{figure}
\centerline{\epsfxsize=3.25in \epsfbox{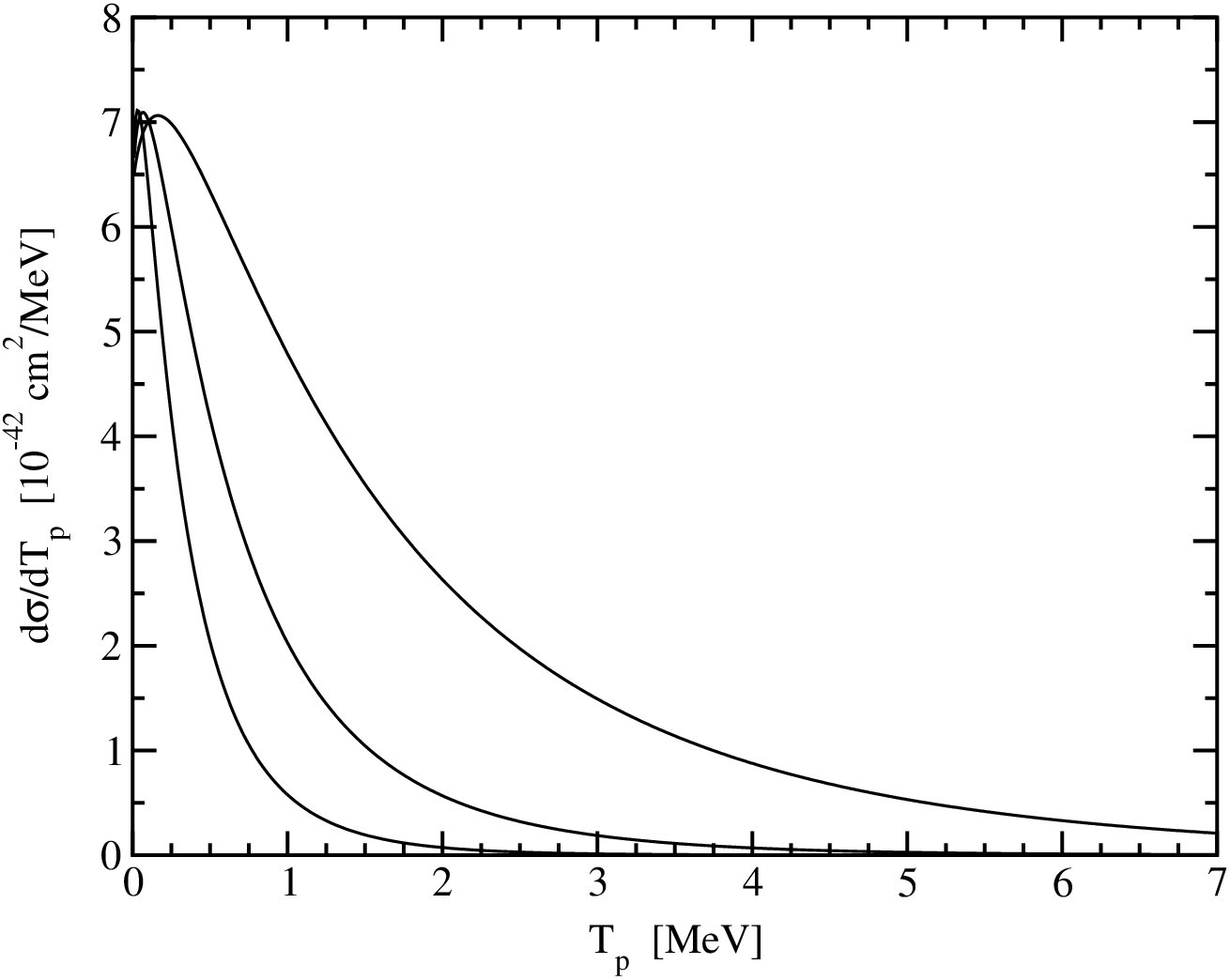}}
\caption{\label{fig:dsigdtavg} The thermally-averaged differential
cross section for Fermi-Dirac distributions of temperature $T = 3.5,
5, 8$ MeV, from left to right.   This illustrates how the proton spectrum
changes with the assumed neutrino temperature (since this is a
neutral-current cross section, it is flavor-independent).}
\end{figure}
%%%%%%%%%%%%%%%%%%%%%

\section{Proton Recoil Spectrum}

The elastically-scattered protons will have kinetic energies
of a few MeV.  Obviously, these very nonrelativistic protons will be
completely invisible in any \v{C}erenkov detector like
Super-Kamiokande.  However, such small energy depositions can be
readily detected in scintillator detectors such as KamLAND and
Borexino.  We first consider the true proton spectrum, and then in the
next Section, we consider how this spectrum would appear in a
realistic detector.

%%%%%%%%%%%%%%%%%%%%
\begin{figure}
\centerline{\epsfxsize=3.25in \epsfbox{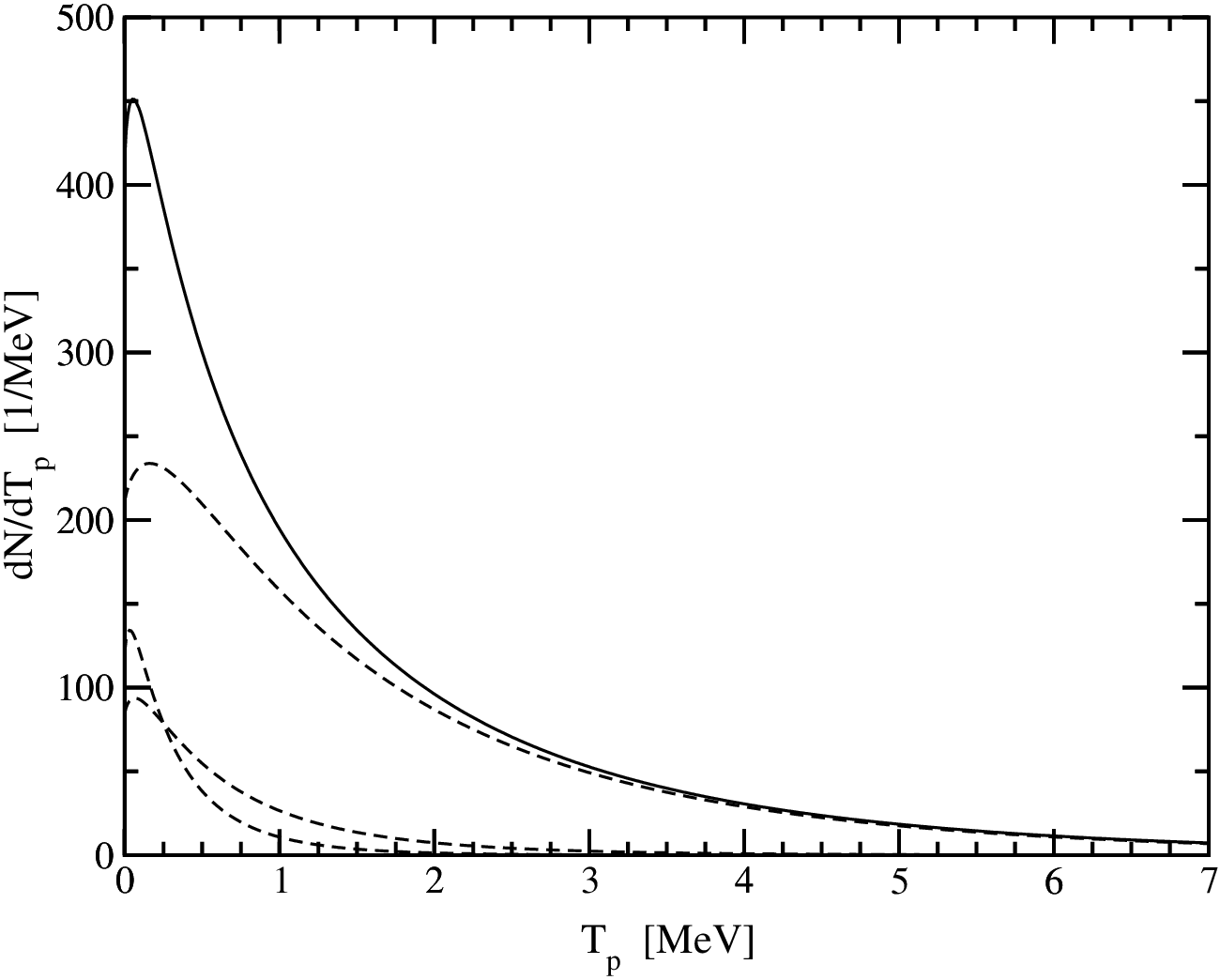}}
\caption{\label{fig:dndt} The true proton spectrum in KamLAND, for a
standard supernova at 10 kpc.  In order of increasing maximum kinetic
energy, the contributions from $\nu_e$, $\bar{\nu}_e$, and the sum of
$\nu_\mu$, $\nu_{\tau}$, $\bar{\nu}_{\mu}$, and $\bar{\nu}_{\tau}$ are
shown with dashed lines.  The solid line is the sum spectrum for all
flavors.  Taking the detector properties into account substantially
modifies these results, as shown below.}
\end{figure}
%%%%%%%%%%%%%%%%%%%%

The true proton spectrum (for one flavor of neutrino) is given by
\begin{equation}
\label{pspec}
\frac{dN}{dT_p}\left(T_p\right) = 
C \int_{(E_\nu)_{min}}^{\infty}
{dE_\nu\, f(E_\nu) \, \frac{d\sigma}{dT_p}\left(E_\nu, T_p\right)}\,,
\end{equation}
where $f(E_\nu)$ is a normalized Fermi-Dirac spectrum and the
differential cross section is given by Eq.~(\ref{diffxs}).  For a
given $T_p$, the minimum required neutrino energy is
\begin{equation}
(E_\nu)_{min} = \frac{T_p + \sqrt{T_p (T_p + 2 M_p)}}{2} 
\simeq \sqrt{\frac{M_p T_p}{2}} \,.
\end{equation}
The normalization constant $C$ is determined by Eq.~(\ref{yield}),
as the integral of Eq.~(\ref{pspec}) over all $T_p$ without the $C$ 
factor is $\langle \sigma \rangle$.

In Fig.~\ref{fig:dsigdtavg}, we show $d\sigma/dT_p$ weighted by
normalized Fermi-Dirac distributions of various temperatures, for a
single neutrino flavor.  Throughout this paper, we refer to the
$\nu_e$ ($T = 3.5$ MeV), $\bar{\nu}_e$ ($T = 5$ MeV), and the combined
$\nu_\mu$, $\nu_{\tau}$, $\bar{\nu}_{\mu}$, and $\bar{\nu}_{\tau}$ ($T
= 8$ MeV) flavors.  Since we know that there are neutrino
oscillations, this language is somewhat incorrect.  However, our
results are {\it totally insensitive} to any oscillations among
active neutrinos or antineutrinos (since this is a neutral-current 
cross section), and also to oscillations between active neutrinos
and antineutrinos (since the cross section is dominated by the
$c_A^2$ terms).
Thus when we refer to the $\nu_e$ flavor, we mean ``those
neutrinos emitted with a temperature $T = 3.5$ MeV, whatever their
flavor composition now,'' etc.

The true proton spectra corresponding to the various flavors are shown
in Fig.~\ref{fig:dndt}.  As seen in the figure, the contributions of
$\nu_e$ and $\bar{\nu}_e$ are quite suppressed relative to the sum of
$\nu_\mu$, $\nu_{\tau}$, $\bar{\nu}_{\mu}$, and $\bar{\nu}_{\tau}$.

%%%%%%%%%%%%%%%%%%%%%%%%%%%%%%%%%%%%%%%%%%%%%%%%%%%%%%%%%%%%%%%%%%%%%%%%%%%%%
%				Quenching				    %
%%%%%%%%%%%%%%%%%%%%%%%%%%%%%%%%%%%%%%%%%%%%%%%%%%%%%%%%%%%%%%%%%%%%%%%%%%%%%

\section{quenching}
\label{quenching}

Low-energy protons lose energy very quickly by ionization.  The energy
loss rate $dE/dx$ of nonrelativistic particles scales roughly as
$dE/dx \sim - z^2/\beta^2$ in this energy range~\cite{RPP}, 
where $z$ is the particle charge and
$\beta$ its velocity.  In contrast to the usual $- 2$ MeV/g/cm$^2$ for a
minimum-ionizing particle, for few-MeV protons, $dE/dx \sim - 100$
MeV/g/cm$^2$.  Thus even a 10 MeV proton will be brought to rest in
less than about 0.1 cm.  In contrast, the hadronic interaction length
for the proton to scatter from a free or bound nucleon is of order 1
cm or larger.  Thus the hadronic energy losses can be totally
neglected; see also Fig.~23.1 of Ref.~\cite{RPP}.  Because of the
nonlinear response of the detector to proton recoil energies, as we
are about to describe, it is important that the original proton energy
is not shared among two or more protons, i.e., from elastic hadronic
scattering.

In a scintillator, there is generally an efficient transfer between
the ionization loss of a charged particle and the detectable
scintillation light observed by phototubes.  For example, in KamLAND,
there are approximately 200 detected photoelectrons per MeV deposited
for a minimum-ionizing particle like an electron~\cite{kamland}.

However, for highly ionizing particles like low-energy protons, the
light output is reduced or ``quenched'' relative to the light output
for an electron depositing the same amount of energy.  The observable
light output $E_{equiv}$ (i.e., equivalent to an electron of energy
$E_{equiv}$) is given by Birk's Law~\cite{leobook}:
\begin{equation}
\label{birks}
\frac{dE_{equiv}}{dx} = \frac{dE/dx}{1 + k_B (dE/dx)}
\end{equation}
where $k_B$ is a constant of the scintillation material, and $dE/dx$
is the energy deposition rate, now in MeV/cm (and defined to be positive).
We assume $k_B \simeq 0.015$ cm/MeV for KamLAND~\cite{kamland}.  For 
small $dE/dx$, the measured light output of a proton is equivalent to that 
from an electron of the same energy.  But for
$dE/dx \sim 100$ MeV/cm, the two terms in the denominator are
comparable, and the light output is reduced.  At still higher $dE/dx$,
then $dE_{equiv}/dx$ tends to a constant.  Birk's Law can thus
reflects a saturation effect: once $dE/dx$ is large, making it larger
does not increase the light output.  Effectively, if all scintillation
molecules along the path of the particle are already excited, any further 
energy deposition is not converted to visible scintillation light.

%%%%%%%%%%%%%%%%%%%%
\begin{figure}
\centerline{\epsfxsize=3.25in \epsfbox{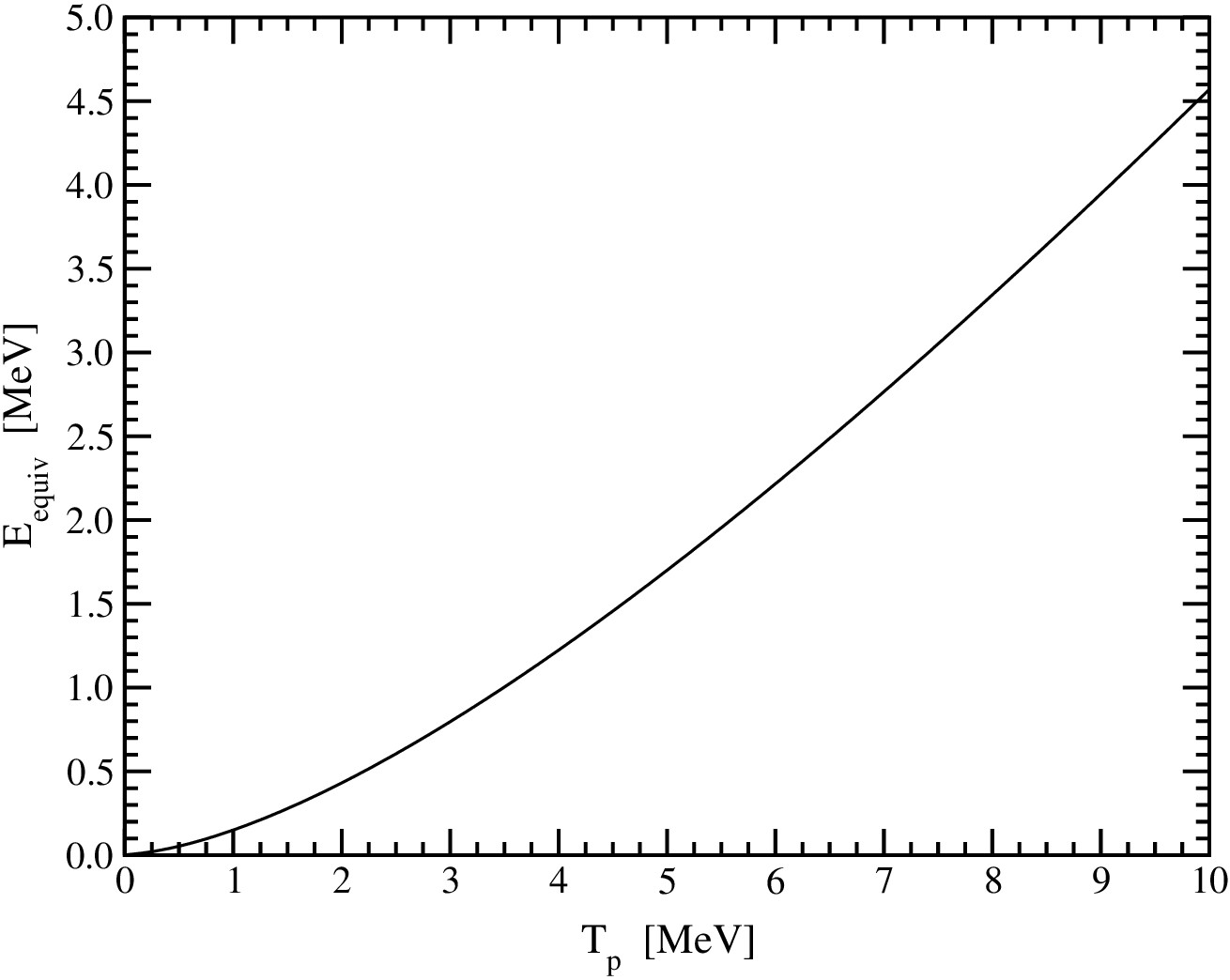}}
\caption{\label{fig:quench} The quenched energy deposit (equivalent
electron energy) as a function of the proton kinetic energy.  The
KamLAND detector properties are assumed.}
\end{figure}
%%%%%%%%%%%%%%%%%%%%

The proton quenching factor was calculated by integrating
Eq.~(\ref{birks}) with tables~\cite{pstar} of $dE/dx$ for protons in
the KamLAND oil-scintillator mixture~\cite{kamland}:
\begin{equation}
\label{equive}
E_{equiv}(T_p) = \int_{0}^{Tp} {\frac{dE}{1 + k_B (dE/dx)}}\,.
\end{equation}
The observed energy in terms of the proton kinetic energy is shown in
Fig.~\ref{fig:quench}.  Thus the proton quenching factor
($E_{equiv}/T_p$) is thus roughly 1/2 at 10 MeV, 1/3 at 6 MeV, 1/4 at 3 MeV,
and so on.  The detector response is nonlinear, though in
well-understood way.  A similar calculation using $\alpha$ particles
recovered the quenching factor of approximately 1/14 noted in
Ref.~\cite{kamland}.  Since the energy deposition scales 
roughly as $dE/dx \sim z^2/\beta^2$, quenching for
alpha particles is much worse than for protons of the same kinetic
energy, since $dE/dx$ is approximately $4 \times 4 = 16$ times larger.
Our results for the proton quenching factor are in good agreement with
direct measurements in a variety of scintillators~\cite{leobook,routine}.

\begin{figure}
\centerline{\epsfxsize=3.25in \epsfbox{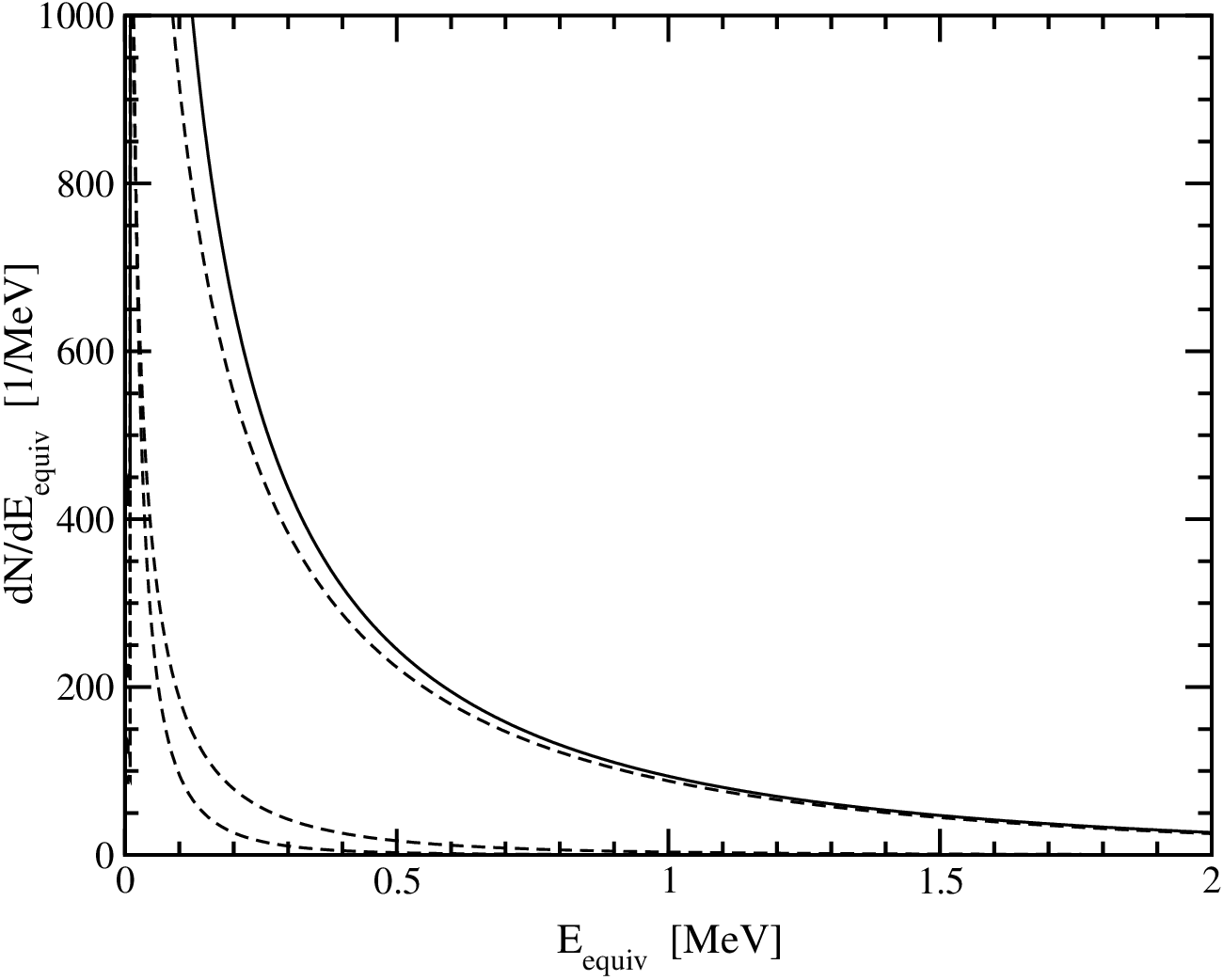}}
\caption{\label{fig:dndtquench} Analogous to Fig.~\ref{fig:dndt};
the struck proton spectrum for the different flavors, but with
quenching effects taken into account.  In order of increasing maximum
kinetic energy, the contributions from $\nu_e$, $\bar{\nu}_e$, and the
sum of $\nu_\mu$, $\nu_{\tau}$, $\bar{\nu}_{\mu}$, and
$\bar{\nu}_{\tau}$ are shown with dashed lines.  The solid line is the
sum spectrum for all flavors.  We assume a 1 kton detector mass for 
KamLAND.}
\end{figure}

Using the quenching function shown in Fig.~\ref{fig:quench}, we can
transform the true proton spectrum shown in Fig.~\ref{fig:dndt} into
the expected measured proton spectrum, shown in
Fig.~\ref{fig:dndtquench}.  If the quenching factor were a constant,
it would simply change the units of the $T_p$ axis.  However, it is
nonlinear, and reduces the light output of the lowest recoils the
most.  This increases the effect, shown in previous figures, that the
measurable contribution from $\nu_e$ and $\bar{\nu}_e$ is highly
suppressed relative to the sum of $\nu_\mu$, $\nu_{\tau}$,
$\bar{\nu}_{\mu}$, and $\bar{\nu}_{\tau}$.

%%%%%%%%%%%%%%%%%%%%%%
%%%%%%%%%%%%%%%%%%%%%%

\begin{figure}
\centerline{\epsfxsize=3.25in \epsfbox{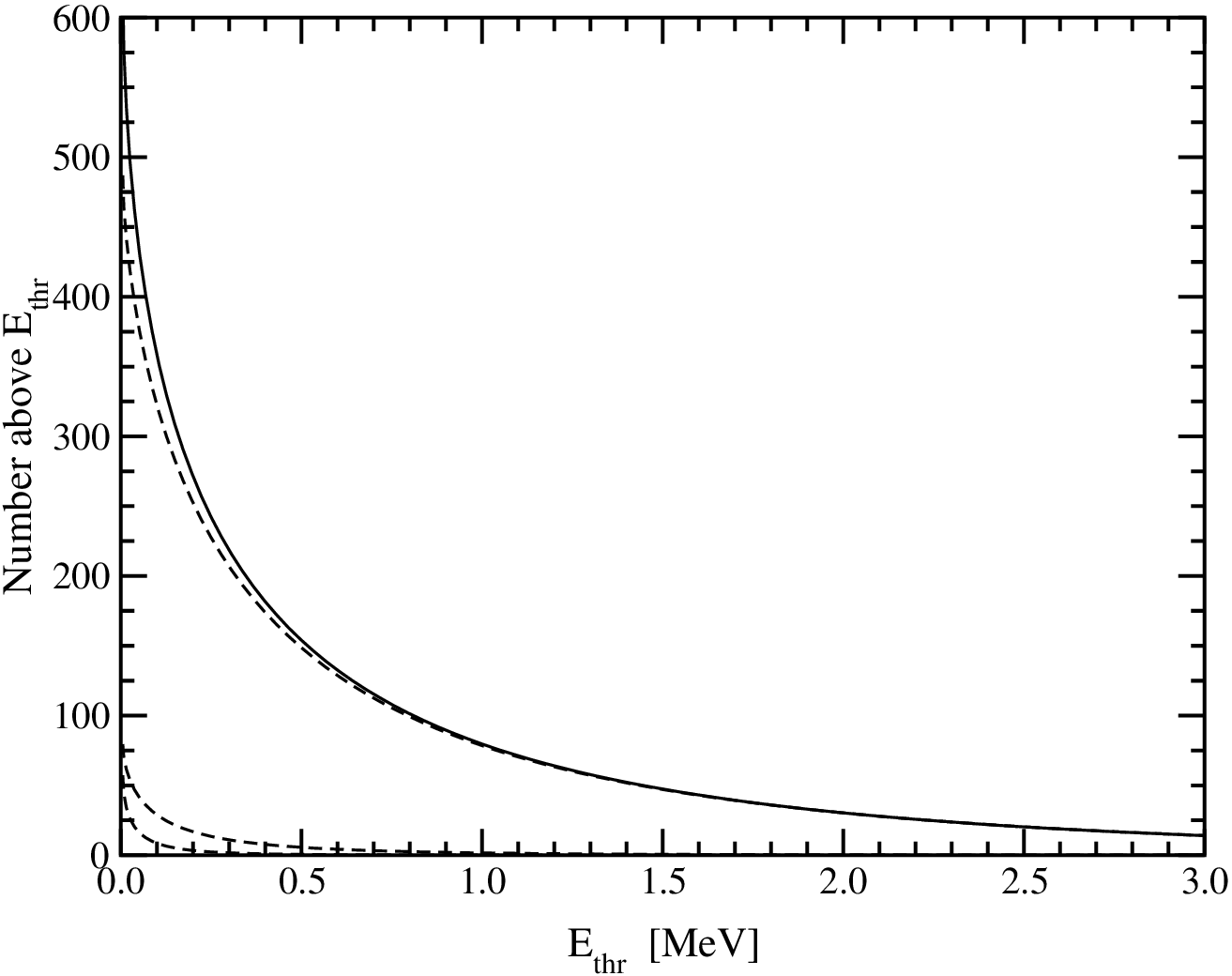}}
\caption{\label{fig:abovethr} The number of events above threshold
in KamLAND as a function of $E_{thr}$ in electron equivalent energy
$E_{equiv}$.
In order of increasing maximum kinetic energy, the contributions from
$\nu_e$, $\bar{\nu}_e$, and the sum of $\nu_\mu$, $\nu_{\tau}$,
$\bar{\nu}_{\mu}$, and $\bar{\nu}_{\tau}$ are shown with dashed lines.
The solid line is for the sum of all flavors.  We assume a
threshold of 0.2 MeV electron equivalent energy in KamLAND.}
\end{figure}

As shown, quenching distorts the spectra according to a known
nonlinear function.  It also reduces the number of events
above threshold.  The anticipated threshold in KamLAND is 0.2 MeV
electron equivalent energy (strictly speaking, KamLAND and Borexino
have somewhat higher target thresholds of approximately 0.28 and 
0.25 MeV, set by background rates; over the short duration of the
supernova pulse, much higher background rates can be tolerated).
With the expected proton quenching, this
corresponds to a threshold on the true proton kinetic energy of 1.2
MeV.  The number of events above this threshold for each flavor
appears in Table~\ref{tab:numevents}.  Above an electron equivalent
threshold of 0.2 MeV, the neutrino-proton elastic scattering yields
from $\nu_e$ and $\bar{\nu}_e$ are quite small.  Thus the measured
proton spectrum will primarily reflect the shape of the underlying
Fermi-Dirac spectrum for the sum of $\nu_{\mu}$, $\nu_{\tau}$
$\bar{\nu}_{\mu}$, and $\bar{\nu}_{\tau}$.  Of course, this has been
convolved with both the differential cross section (which gives a
range of $T_p$ for a given $E_\nu$), and also the effects of
quenching.  However, as we will show, the properties
of the initial neutrino spectrum can still be reliably deduced.  The numbers
of events above a given electron equivalent threshold are shown in
Fig.~\ref{fig:abovethr}.

%%%%%%%%%%%%%%%%%%%%%%
\begin{table}[b]
\caption{Numbers of events in KamLAND (1 kton mass assumed)
above the noted thresholds for a
standard supernova at 10 kpc, for the separate flavors or their
equivalents after oscillations.
Oscillations do not change the number of neutrinos at a
given energy, and the neutral-current yields are insensitive to the
neutrino flavor.  Equipartition among the six flavors is assumed (see
the text for discussion).  The thresholds are in electron equivalent
energy, and correspond to minimum true proton kinetic energies of 0
and 1.2 MeV.  As discussed in Section~\ref{cross}, weak magnetism
corrections are not included.}
\begin{tabular}{l|rcr}
\it{Neutrino Spectrum} & $E_{thr} = 0$ & \phantom{xxx} & $ 0.2$ MeV \\
\hline\hline
$\nu: T = 3.5$ MeV			& 57 	&& 3 	\\ \hline 
$\bar{\nu}: T = 5$ MeV			& 80 	&& 17 	\\ \hline 
$2 \nu: T = 8$ MeV			& 244	&& 127	\\ \hline 
$2 \bar{\nu}: T = 8$ MeV	 	& 243	&& 126	\\ \hline 
All 					& 624	&& 273	\\
\hline\hline
\end{tabular}
\label{tab:numevents}
\end{table}
%%%%%%%%%%%%%%%%%%%%%%

%%%%%%%%%%%%%%%%%%%%%%
%%%%%%%%%%%%%%%%%%%%%%

%%%%%%%%%%%%%%%%%%%%%%%%%%%%%%%%%%%%%%%%%%%%%%%%%%%%%%%%%%%%%%%%%%%%%%%%%%%%%
%				Backgrounds				    %
%%%%%%%%%%%%%%%%%%%%%%%%%%%%%%%%%%%%%%%%%%%%%%%%%%%%%%%%%%%%%%%%%%%%%%%%%%%%%

\section{Backgrounds}

In this Section, we consider several backgrounds to the signal of
neutrino-proton elastic scattering from supernova $\nu_\mu$, $\nu_{\tau}$,
$\bar{\nu}_{\mu}$, and $\bar{\nu}_{\tau}$.

The first is neutrino-proton elastic scattering from $\nu_e$ and
$\bar{\nu}_e$.  As shown above, in particular in
Figs.~\ref{fig:dndtquench} and \ref{fig:abovethr}, this
contribution is minimal above the expected threshold.  We assume that
it can be statistically subtracted using knowledge of $\nu_e$ and
$\bar{\nu}_e$ temperatures measured in charged-current reactions
and do not consider it further.

The second comes from a variety of other charged-current supernova
neutrino signals in the detector.  As noted above, approximately 310
events are expected from $\bar{\nu}_e + p \rightarrow e^+ +
n$~\cite{invbeta}.  These events can easily be identified by the tight
coincidence (roughly a few times 10 cm in position, and 0.2 ms in time) 
in the detection
of the high-energy (about 20 MeV) positrons and the subsequent neutron
captures on protons (2.2 MeV gamma).  There are also charged-current
reactions of $\nu_e$ and $\bar{\nu}_e$ on $^{12}$C, proceeding almost
exclusively to the ground states of $^{12}$N and $^{12}$B,
respectively.  About 10 events are expected, and possibly a few times
more if oscillations effectively swap spectra~\cite{kamland}.
However, these events can be identified by the subsequent $^{12}$N and
$^{12}$B beta decays, with lifetimes of order 10 ms and electron
endpoints of order 15 MeV.  The total yield from neutrino-electron
elastic scattering (technically, mixed charged- and neutral-current)
is expected to be about 20 events.  We assume that these events can be
statistically subtracted from the spectrum, or that particle
identification by pulse-shape-discrimination (PSD) will be possible.

The third comes from other neutral-current supernova neutrino signals
in the detector.  The best-known is the superallowed neutral-current
excitation of the 15.11 MeV state in $^{12}$C, which decays by gamma
emission.  About 60 events are expected, and they will be easily
identified by their narrow spectrum at 15.11 MeV~\cite{kamland}.
There are also inelastic neutral-current excitations of $^{12}$C
that decay by
particle emission.  The yield from all channels that emit a proton is
about 45 events, using the cross sections and branching ratios of
Ref.~\cite{nuex}.  Some fraction, probably most, will be above
threshold and will add to the signal of low-energy protons.  Though
the proton spectra for these reactions have not been published, we
assume that their contribution will be known and can be subtracted.
The yield from all channels that emit a neutron is about 20 events,
again using the results of Ref.~\cite{nuex}.  These neutrons will be
captured, giving 2.2 MeV gamma rays.  These inelastic neutral-current
reactions
with proton and neutron emission have not previously been recognized
as supernova neutrino detection channels.  Finally, since a 50 MeV
neutrino corresponds to a wavelength of 4 fm, about the diameter of a
carbon nucleus, there can also be coherent neutral-current scattering
of the whole nucleus~\cite{coherent}.  The number of events is very
large, coincidentally as large as the total neutrino-proton elastic
scattering yield (neglecting the detector threshold).
However, the expected recoil kinetic energies are of course
about 12 times smaller than for free protons.  Additionally, since it
is spin-independent vector scattering, the {\it smallest} recoil
energies are favored.  (In contrast, neutrino-proton elastic
scattering is spin-dependent, and the proton spin is flipped in the
scattering.)  The recoil carbon ions will be very heavily quenched,
and so this signal is unobservable in a detector like KamLAND.

The fourth comes from cosmic-ray induced detector backgrounds.
Because it is located deep underground, the muon rate in KamLAND
scintillator is only about 0.3 Hz, and so all muon-related backgrounds
are very small over the short duration of the supernova
burst~\cite{kamland}.

The fifth and most serious comes from low-energy radioactivities in
and around the detector.  Normally, these are not a concern for
relatively high-energy supernova events.  However, here we are
considering signals down to about 0.2 MeV detected energy, where many
different radioactive backgrounds contribute.  At present, KamLAND is
configured to detect few-MeV reactor antineutrinos via the
coincidence between the prompt positrons and the delayed neutron
captures, and low singles backgrounds above 0.2 MeV are not required.
Published data on the KamLAND background spectrum are not yet
available.  However, if KamLAND is to be eventually used for detecting $^7$Be 
solar neutrinos by neutrino-electron scattering, then the background
in this energy range will have to be reduced to about $10^{-3}$ Hz,
the rate of solar neutrino events expected (similar considerations
hold for Borexino).  For the supernova signal
discussed in this paper, a much larger background rate of about 1 Hz
could be tolerated.  This rate is set by the consideration of
being much less than (300 events/10 s) = 30 Hz.

Therefore, in what follows we consider just the main signal, and
neglect backgrounds.

%%%%%%%%%%%%%%%%%%%%%%%%%%%%%%%%%%%%%%%%%%%%%%%%%%%%%%%%%%%%%%%%%%%%%%%%%%%%%
%				Fits					    %
%%%%%%%%%%%%%%%%%%%%%%%%%%%%%%%%%%%%%%%%%%%%%%%%%%%%%%%%%%%%%%%%%%%%%%%%%%%%%

\section{Proton Spectrum Fits}
\label{fits}

In this Section, we show how the measured proton spectrum can be used
to separately determine the total energy of the $\nu_\mu$, $\nu_{\tau}$,
$\bar{\nu}_{\mu}$, and $\bar{\nu}_{\tau}$ neutrinos {\it and} their
time-averaged temperature.
The total number of detected events is proportional to the portion of
the total binding energy carried away by these four flavors, and we
denote this by $E^{tot}$ (note that this is {\it not} the total
binding energy $E_B$).  For a standard supernova, 
$E_{tot} = 4 (E_B/6) = 2/3 E_B \simeq 2
\times 10^{53}$ ergs.  We denote the temperature of these four flavors
by $T$.  If only the total yield were measured, as for most
neutral-current reactions, there would be an unresolved degeneracy
between $E^{tot}$ and $T$, since
\begin{equation}
N \sim E^{tot} \frac{\langle \sigma \rangle}{T}\,.
\end{equation}
Note that for $\sigma \sim E_\nu^n$, then $\langle \sigma \rangle \sim
T^n$.  For $\nu + d \rightarrow \nu + p + n$ in SNO, for example,
$\sigma \sim E^2$, so $N \sim E^{tot} T$.  Thus for a given
measured number of events, one would only be able to define a
hyperbola in the plane of $E^{tot}$ and $T$.  The scaling is
less simple here because of threshold effects, but the idea is the
same.

Here we have crucial information on the shape of the neutrino
spectrum, revealed through the proton spectrum.  To remind the reader,
in most neutral-current reactions there is {\it no} information on the
neutrino energy, e.g., one only counts the numbers of thermalized
neutron captures, or measures nuclear gamma rays (the energies of
which depend only on nuclear level splittings).

In this Section, we perform quantitative tests of how well the
parameters $E^{tot}$ and $T$ can be determined from the
measured proton spectrum.  (We did also investigate the effects of a
chemical potential in the Fermi-Dirac distribution, but found that it
had little effect.  This is simply because the cross section is not
rising quickly enough to see the tail of the thermal distribution in
detail~\cite{SNsno}.)  Of course, if the distance to the supernova is
not known, then we are effectively fitting for $E^{tot}/D^2$.

\begin{figure}
\centerline{\epsfxsize=3.25in \epsfbox{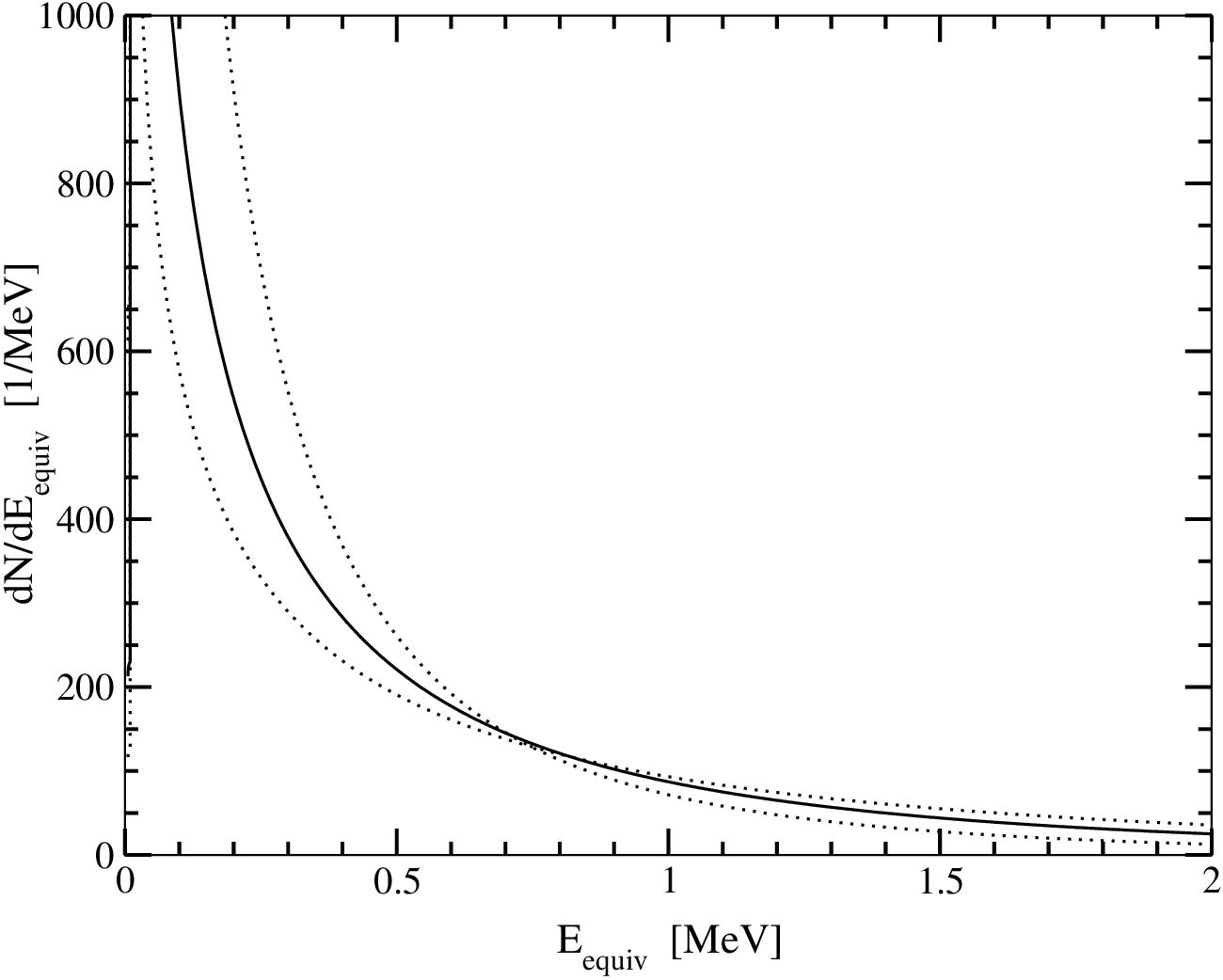}}
\caption{\label{fig:varspecs} Example spectra with different values of
$E^{tot}$ and $T$, all chosen to give the {\it same} number
of events above an electron equivalent threshold of 0.2 MeV (true
proton energy 1.2 MeV) in KamLAND.  Though not shown in this figure,
the spectrum above 2 MeV is included in our analysis.
At the 0.2 MeV point, from left to right
these correspond to $(E^{tot}, T)$ = (4.2, 6), (2.0, 8),
(1.4, 10), respectively, with $E^{tot}$ in $10^{53}$ ergs and
$T$ in MeV.  (Thus with the standard $E_{tot} = 2 \times 10^{53}$ ergs,
the number of events above threshold with $T = 6$ MeV is $2.0/4.2$
times the number with $T = 8$ MeV; with $T = 10$ MeV, it is $2.0/1.4$
times the number with $T = 8$ MeV).}.
\end{figure}

We performed Monte Carlo simulations of the supernova signal in
KamLAND and made chi-squared fits to determine $E^{tot}$ and
$T$ for each fake supernova.
To perform the fits, we started with an ``ideal'' spectrum, as
described by the integral:
\newpage
\begin{eqnarray}
\label{idealspec}
\left(\frac{dN}{dT_p}\right)_{{\rm ideal}} & = &
C \int_0^\infty dT'_p\, G(T'_p; T_p, \delta T_p) \nonumber \\
& \times & \int_{(E_{\nu})_{min}}^\infty \hspace{-0.5cm}
dE_\nu\, f(E_\nu)\, \frac{d\sigma}{dT'_p}(E_\nu, T'_p)\,,
\end{eqnarray}
where the inner integral is as in Eq.~(\ref{pspec}), with the addition
that quenching corrections are applied to $T'_p$ after convolution
with $f(E_\nu)$.  For the Gaussian
energy resolution $G(T'_p; T_p, \delta T_p)$, we used $\delta T_p =
0.1\sqrt{T_p/(1\ {\rm MeV})}$~\cite{kamland}.  Because the proton
spectrum has already been smeared by the neutrino spectrum and the
differential cross section, the Gaussian energy resolution has only a
minor effect.  The normalization constant $C$ is given by comparison
to Eq.~(\ref{yield}).  Example spectra are shown in
Fig.~\ref{fig:varspecs}.

%%%%%%%%%%%%%%%%%%%%%%
%%%%%%%%%%%%%%%%%%%%%%

Using $(dN/dT_p)_{\rm ideal}$, we binned the spectrum by the
following integral:
\begin{equation}
\label{binning}
N_i = \int_{(E_{min})_i}^{(E_{max})_i} dT_p
	\left(\frac{dN}{dT_p}\right)_{\rm ideal}
\end{equation}
where $N_i$ is the number of events in bin $i$, and $(E_{min})_i$ and
$(E_{max})_i$ are the minimum and maximum energies for bin $i$.
Eight bins of variable width were used, chosen to contain roughly
the same number of expected events per bin.
For a chosen $E^{tot}$ and $T$, this was the starting point
of our Monte Carlo (and the bin boundaries were kept fixed).
For each fake supernova, we sampled the number of
events in each of these bins according to the appropriate Poisson
distributions.  The resulting spectrum was as one might obtain from a
single supernova, given the finite number of events expected.  We then
varied $E^{tot}$ and $T$ in Eq.~(\ref{binning}) until the
values that best fit the fake spectral data were determined.  For a
given set of assumed $E^{tot}$ and $T$, this procedure was
repeated many times.  The distributions of the final
$E^{tot}$ and $T$ thus reveal the expected errors on fitting
$E^{tot}$ and $T$ for a single real future supernova.

\begin{figure}
\centerline{\epsfxsize=3.25in \epsfbox{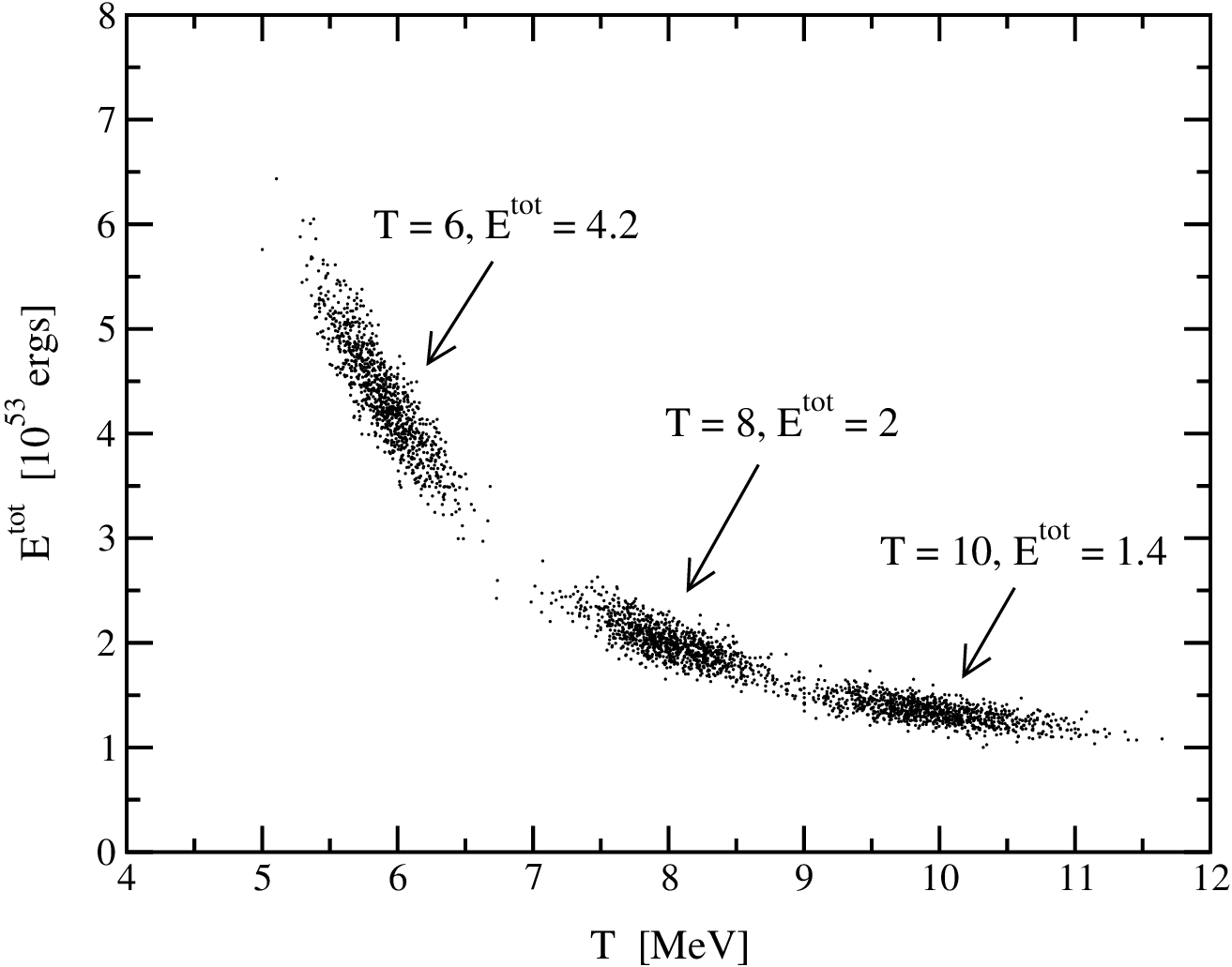}}
\caption{\label{fig:fits} Scatterplot of $10^3$ fitted values, in the
$E^{tot}$ and $T$ plane, for the labeled ``true'' values,
where $E^{tot}$ is the total portion of the binding energy
carried away by the sum of $\nu_\mu$, $\nu_{\tau}$, $\bar{\nu}_{\mu}$,
and $\bar{\nu}_{\tau}$, and $T$ is their temperature.  The values of
$E^{tot}$ and $T$ were chosen such that the numbers of
events above threshold were the same.  The measured shape of the
proton spectrum breaks the degeneracy between these two parameters.
Without that spectral information, one could not distinguish between
combinations of $E^{tot}$ and $T$ along the band in this
plane that our three example regions lie along.}
\end{figure}

Three examples are shown in Fig.~\ref{fig:fits}, where one can see
that $E^{tot}$ and $T$ can each be determined with
roughly 10\% error.  These errors scale as $1/\sqrt{N}$, where $N$ is
the total number of events (i.e., if one imagines a detector of a
different mass or a different assumed supernova distance).  If the
distance were completely uncertain, one would not be able to determine
$E^{tot}$.  However, after marginalizing over the unknown
$E^{tot}$ (i.e., projecting these scatterplots onto the $T$
axis), one would still obtain a good measurement of $T$.

%%%%%%%%%%%%%%%%%%%%
%%%%%%%%%%%%%%%%%%%%%

%%%%%%%%%%%%%%%%%%%%%%%%%%%%%%%%%%%%%%%%%%%%%%%%%%%%%%%%%%%%%%%%%%%%%%%%%%%%%
%			Conclusions					    %
%%%%%%%%%%%%%%%%%%%%%%%%%%%%%%%%%%%%%%%%%%%%%%%%%%%%%%%%%%%%%%%%%%%%%%%%%%%%%

\section{Discussion and Conclusions}

We have shown that neutrino-proton elastic scattering, previously
unrecognized as a useful detection reaction for low-energy neutrinos,
in fact has a yield for a supernova comparable to $\bar{\nu}_e + p
\rightarrow e^+ + n$, even after taking into account the quenching of
the proton scintillation light and assuming a realistic detector
threshold.

In addition, the measured proton spectrum is related to
the incident neutrino spectrum.  We have shown explicitly that one can
separately measure the total energy and temperature of
$\nu_\mu$, $\nu_\tau$, $\bar{\nu}_\mu$, and $\bar{\nu}_\tau$, each
with uncertainty of order 10\% in KamLAND.  This greatly enhances the 
importance of detectors like KamLAND and Borexino for detecting supernova
neutrinos.

For Borexino, the useful volume for supernova neutrinos is 0.3 kton,
and the hydrogen to carbon ratio in the pure pseudocumene
(C$_9$H$_{12}$) is $1.3:1$~\cite{SNborexino}, so there are about 4.7
times fewer free proton targets than assumed for KamLAND.  However,
the quenching is less in pure scintillator (KamLAND is about 20\%
pseudocumene and 80\% paraffin oil~\cite{kamland}), and the errors on
$E^{tot}$ and $T$ scale as $1/\sqrt{N}$, so that the precision in
Borexino should be about 20\% or better.

Other techniques for bolometric measurements of supernova neutrino
fluxes have been studied.  Detectors for elastic neutral-current
neutrino scattering on electrons~\cite{cabrera} and coherently on
whole nuclei~\cite{coherent} have been discussed, but never built.
If neutrino oscillations are effective in swapping spectra, then the
temperature of the ``hot'' flavors may be revealed in the measured
positron spectrum from $\bar{\nu}_e + p \rightarrow e^+ + n$; two
recent studies have shown very good precision ($\lesssim 5\%$) for
measuring the temperatures and the total binding
energy~\cite{barger,minakata}.  However, they assumed exact energy
equipartition among the six neutrino flavors, whereas the uncertainty
on equipartition is at least 50\%~\cite{raffeltproc}.  Nevertheless,
under less restrictive assumptions, this technique may play a
complementary role.  Finally, since for different cross sections, the
neutral-current yields depend differently on temperature, comparison
of the yields may provide some information~\cite{relyields}.  However,
there are caveats.  In neutrino-electron scattering, the neutrino
energy is not measured because the neutrino-electron angle is much
less than the angular resolution due to multiple scattering.  The
scattered electrons, even those in a forward cone, sit on a much
larger background of $\bar{\nu}_e + p \rightarrow e^+ + n$ events, so
it is difficult to measure their spectrum~\cite{SNpoint}; also, their
total yield is only weakly dependent on temperature.  At the other
extreme (see Fig.~3 of Ref.~\cite{relyields}), the yield of
neutral-current events~\cite{SNsk} on $^{16}$O depends strongly on a
possible chemical potential term in the thermal distribution.

It is important to note that the detection of recoil protons from {\it
neutron}-proton elastic scattering at several MeV has been routinely
accomplished in scintillator detectors (see, e.g.,
Ref.~\cite{routine}).  Since both particles are massive, the proton
will typically take half of the neutron energy.  This reaction
provides protons in the same energy range as those struck in
neutrino-proton elastic scattering with $E_\nu \sim 30$ MeV.  This is
a very important proof of concept for all aspects of the detection of
low-energy protons.

Though low-energy backgrounds will be challenging, it is also
important to note that the background requirements for detecting the
supernova signal are approximately 3 orders of magnitude {\it less}
stringent than those required for detecting solar neutrinos in the
same energy range (taking quenching into account for our signal).
Borexino has been designed to detect very low-energy solar neutrinos,
and KamLAND hopes to do so in a later phase of the experiment.

These measurements would be considered in combination with similar
measurements for $\nu_e$ and $\bar{\nu}_e$ from charged-current
reactions in other detectors.  Separate measurements of the
total energy and temperature for each flavor will be
invaluable for comparing to numerical supernova models.  They will
also be required to make model-independent studies of the effects of
neutrino oscillations.  If the total energy release $E_B$ in all
flavors has been measured, then 
\begin{equation}
E_B \simeq \frac{3}{5} \frac{G M_{NS}^2}{R_{NS}}\,,
\end{equation}
thus allowing a direct and unique measurement of the newly-formed 
neutron star properties, principally the mass $M_{NS}$~\cite{NS}.

%%%%%%%%%%%%%%%%%%%%%%%%%%%%%%%%%%%%%%%%%%%%%%%%%%%%%%%%%%%%%%%%%%%%%%%
%			Acknowledgements			      %
%%%%%%%%%%%%%%%%%%%%%%%%%%%%%%%%%%%%%%%%%%%%%%%%%%%%%%%%%%%%%%%%%%%%%%%

\section*{ACKNOWLEDGEMENTS}

We thank Felix Boehm, Laura Cadonati, Frank Calaprice, Mark Chen,
Chuck Horowitz, Thomas Janka, Glenn Horton-Smith, Loren Hoffman, Bob
McKeown, Marianne Neff, Lothar Oberauer, Stephen Parke, Andreas
Piepke, Georg Raffelt, Junpei Shirai, Fumihiko Suekane, Rex Tayloe,
Bryan Tipton, and Bruce Vogelaar for discussions.

JFB was supported as a Sherman Fairchild Fellow at Caltech during the
initial part of this project, and as the David N. Schramm Fellow at
Fermilab during the final part.  Fermilab is operated by URA under DOE
contract No. DE-AC02-76CH03000.  JFB was additionally supported by
NASA under NAG5-10842.  WMF was supported by a Dr. and Mrs. Lew Allen,
Jr. Summer Undergraduate Research Fellowship (SURF) at Caltech.  This
work was supported in part by the U.S. Department of Energy under
Grant No. DE-FG03-88ER40397 at Caltech.

%%%%%%%%%%%%%%%%%%%%%%%%%%%%%%%%%%%%%%%%%%%%%%%%%%%%%%%%%%%%%%%%%%%%%%%%%%%%%%
%			Bibliography					     %
%%%%%%%%%%%%%%%%%%%%%%%%%%%%%%%%%%%%%%%%%%%%%%%%%%%%%%%%%%%%%%%%%%%%%%%%%%%%%%

\end{document}